# Fisher's ideas and the design of field experiments in agronomy and plant breeding

Hans-Peter Piepho[1]

[1] Biostatistics Unit, Institute of Crop Science, University of Hohenheim, Stuttgart, Germany

**Abstract**

R. A. Fisher was one of the greatest scientists of the last century. He made many ground-breaking contributions, so many indeed that it seems almost impossible to list all of them. His revolutionary contributions to the design of experiments can mostly be traced back to the early part of his academic career, and they are inextricably linked to his involvement with agricultural field experiments at Rothamsted Experiment Station. In this talk I will review Fisher's key ideas on experimental design and relate them to some of the work I am involved in, most of which directly focuses on field experiments in agriculture. Topics covered include systematic designs, row-column designs, augmented row-column designs, multi-environment trials, partially replicated designs, optimal allocation of trials to zones in sub-divided target populations of environments, and the connection of trialling systems across countries.

## 1 Introduction

It is truly a great honour to be invited to give the 43rd Fisher Memorial Lecture. Just going through the list of eminent people who gave this lecture before was dizzying. Even more breath-taking is the list of R. A. Fisher's ground-breaking achievements in statistics and genetics.

Among the many outstanding contributions made by Fisher (Bodmer et al., 2021), this talk will focus on the design and analysis of experiments. Most of this work can be related to Fisher's early days at Rothamsted Experiment Station, where he was hired as a statistician in 1919 (Fisher Box, 1978). The very first paper introducing the idea of analysis of variance (ANOVA) of field experiments was published in 1923 (Fisher and Mackenzie, 1923). The next major milestone was the publication of his book *Statistical Methods for Research Workers* (SMRW) in 1925 (Fisher, 1925), where the key principles of replication, randomization and local control (blocking) were first laid down for a broader audience. A year thereafter, almost exactly a century ago at the time of this lecture, this was followed by the famous paper entitled "The arrangement of field experiments" (Fisher, 1926), which is perhaps the clearest and most lucid discussion of his principles (Speed, 1991). According to Speed (1991), the publication of the second book entitled *Design of Experiments* (DOE) in 1935 (Fisher, 1935) "signalled the end, not the beginning of the Fisherian revolution in field experimentation." I will subsequently make frequent reference to Fisher's two seminal books (SMRW and DOE). The editions used throughout for specific citations are those edited by J. H. Bennett in 1990 in a single volume entitled "A Re-issue of *Statistical Methods for Research Workers*, *The Design of Experiments* and *Statistical Methods and Scientific Inference*." This volume comprises the Fourteenth Edition of SMRW (Fisher, 1973) and the Eighth Edition of DOE (Fisher, 1971).

Fisher's influence reaches far beyond agriculture of course. At the same time, his revolutionary contributions to the design and analysis of experiments clearly originated in the context of agricultural field experimentation (Rosenberger, 2026). Re-reading Fisher in preparation for this lecture has been a most refreshing experience, especially for someone whose own work focuses almost entirely on applications in agriculture, and field experiments in particular. It is in that context that I would like to review Fisher's key ideas on experimental design and relate them to some of the work I am involved in. The presentation is organized into two major sections, the first on individual trials and the second on multi-environment trials (MET).

## 2 Individual trials

### 2.1 Systematic vs. randomized designs

The first time randomization was explicitly introduced as a device ensuring a valid estimate of error was Fisher (1925). The principle is folklore in any textbook touching on the design of experiments. However, whereas the process or randomization itself is easy to understand, the connection to subsequent analyses is still a topic that is difficult to grasp for many students. Speed (1991) attributes this difficulty to "the absence of any direct connection between the linear models according to which such experiments are usually analysed, and the randomization argument." Up to the point of his writing in 1925, Fisher had not been involved in an experiment designed and analysed according to his proposed principles (Speed, 1991). His exposition was mainly based on analyses of experiments designed by others and on numerical experiments performed based on the uniformity trials of Mercer and Hall (1911). On the face of it, this might raise the impression that the use of uniformity trial data in his 1925 book was just a stop-gap, but I think nothing could be further from the truth. As demonstrated by Fisher, uniformity trials have great scientific value in their own right, because they allow assessing the validity of experimental designs and their associated estimates of error. I also believe they can be very useful didactical tools in teaching experimental design, especially regarding the connection of the randomization argument with the linear model used for analysis.

Systematic designs, in which the arrangement of treatments follows some systematic, unrandomized pattern, were strongly opposed by Fisher. However, they are still used in agricultural field experiments today. One case in kind is on-farm experiments (Piepho et al., 2011; GRDC, 2021; Alesso et al., 2021; Cao et al., 2024). Another pertinent example is long-term experiments, some of which were designed using systematic arrangements. In a collaboration with agronomists I have recently worked with data from such trials in Germany (Macholdt et al., 2019a,b). The systematic designs used in these trials go back to a small book published by the German agronomist and soil scientist Eilhard Alfred Mitscherlich (Mitscherlich 1919, 1925). It is interesting to note that the first edition of Mitscherlich's book was published in 1919, coinciding with the year Fisher took up his position at Rothamsted. Moreover, the year the second edition came out (1925) coincides with the publication year of Fisher's first book (SMRW).

Mitscherlich suggested a method of analysis for a systematic design, which may be illustrated for the case of five treatments A-E with four replicates arranged in a single row of plots in the order ABCDE ABCDE ABCDE ABCDE. The method of analysis suggested for this trial is based on the idea of a sliding block, explained in more detail in Piepho and Vo-Thanh (2020) and Piepho et al. (2026). Briefly, the idea is that a sliding window of five plots can be moved across the row of plots, each time covering five plots of the layout. Each time, the window represents a complete block. Thus, the average of yields on the five plots can be used to make corrections for the observed yield, either only for the central plot (von Lochow and Schuster, 1961), or for all plots in the sliding window (Mitscherlich, 1919, 1925). In keeping with the idea of sliding blocks, one can assume a linear model with block effects for each of the sliding blocks. If it is further assumed that the effects associated with the sliding blocks are independently and identically distributed with constant variance, a spatial covariance model results in which the spatial covariance drops to zero linearly up to a distance equal to the width of the sliding block. Interestingly, if that model is used to generate an A-optimal (or D-optimal) design via computer search, a systematic layout of the form proposed by Mitscherlich results (Piepho and Vo-Thanh, 2020; Piepho et al., 2026). However, each time the design package is re-run, the same systematic layout results, and hence there is obviously no randomization, apart from a permutation of the treatment labels.

The lack of randomization raises the issue of the validity of any estimate of error. This point is at the heart of the famous dispute between Fisher and W. S. Gosset ("Student") about the merits and demerits of systematic vs. randomized designs (Picard, 1980). To demonstrate the validity of the analysis of variance in Fisher's sense, we used the Mercer and Hall (1911) uniformity trial data on wheat, overlaying it with completely randomized designs (CRD) and the randomized complete block designs (RCBD) for five treatments and four replicates (Piepho et al., 2026). Comparing the model-based and empirical standard errors and well as the empirical Type 1 errors, the CRD and RCBD are found to be valid as shown by Fisher. By contrast, using the systematic design by Mitscherlich, use of the uniformity trial data shows that neither the error variance estimate nor the empirical Type 1 error rates are valid. This is also found when the spatial model described in the previous paragraph is used for analysis. These unsurprising findings confirm Fisher's assertion that with systematic arrangements there is no unique method of analysis, and there is also no way of knowing if the analysis is going to be valid. This problem must be weighed against any potential gain in precision a systematic arrangement may entail. In a similar vein, the potential gains in precision that may arise from sophisticated spatial modelling of designed experiments (Cressie and Wikle, 2011, pp 9−10 and § 4; De Faveri et al., 2017) must be balanced against the lack of certainty regarding validity of estimates of error. For this reason, our preference is generally for a randomization-based analysis, considering spatial model components merely as an add-on option (Piepho et al., 2021).

Fisher (1973, p.267) also proposed to use the position number of plots within blocks as a covariate in analysis of covariance (ANCOVA), another one of his ingenious ideas (Cox and McCullagh, 1982). This positional covariate may be related to the systematic design he considers in his introductory example, where he uses the order ABCDE EDCBA ABCDE EDCBA. If plots are numbered 1 to 20 in field order, then the average position number is the same for each treatment. Such a layout may be expected to correct for any linear spatial trends in the layout. It is worth noting that the design is similar to the one proposed by "Student" in what he called the "Sandwich Design", associated with the so-called "Half-Drill Strip Method", involving systematic arrangements of the form AB BA AB BA etc. (Picard, 1980). The systematic design considered by Fisher (1973, p.267) also yields invalid estimates of error if analysed as an RCBD when used on the Mercer and Hall data (Piepho et al., 2026).

A much milder restriction on randomization results if one generates a design assuming a linear model with fixed effects for blocks and treatments, plus a regression on the positional covariate. For the example of five treatments and four replicates, the randomization set comprises 993 designs. Imposing each of these designs in turn onto the Mercer and Hall data, we found empirical Type 1 error rates close to the nominal one and empirical standard errors close to the model-based ones (Piepho et al., 2026), suggesting that such designs provide approximately valid statistical inferences. This is despite the fact that there is no strictly valid randomization theory for ANCOVA (Kempthorne, 1977).

I think this little example, motivated by the exposition in Fisher's two books (Fisher, 1973, §48; Fisher, 1971, §20-21), demonstrates the great utility of uniformity trial data. Conducting uniformity trials seems to have long fallen out of fashion, but I believe such trials are still worth considering and conducting for the purpose of evaluating the empirical performance of new designs, as well as of more sophisticated methods of analysis, including spatial models (Richter and Kroschweski, 2012). Empirical evaluations based on uniformity trial data may also be more convincing to experimenters than assessments based on Monte Carlo simulation, which always raise the question as to the best and most realistic model, if any, to simulate the data. It is so much easier to let uniformity trial data speak for themselves, obviating the need to specify a model for data generation.

Concluding this section, it is worth emphasizing that Fisher viewed the randomization distribution induced by a randomization procedure merely as a tool to verifying validity of the preferred analysis of variance assuming independence and normality of errors. This key point is nicely conveyed in this citation (Fisher, 1971, p.48):

> *In recent years tests using the physical act of randomization to supply (on the Null Hypothesis) a frequency distribution, have been largely advocated under the name of "Non-parametric" tests. Somewhat extravagant claims have often been made on their behalf. The example of this Chapter, published in 1935, was by many years the first of its class. The reader will realise that it was in no sense put forward to supersede the common and expeditious tests based on the Gaussian theory of errors. The utility of such nonparametric tests consists in their being able to supply confirmation whenever, rightly or, more often, wrongly, it is suspected that the simpler tests have been appreciably injured by departures from normality.*

This point is at the heart of Fisher's principle of randomization. It appears to me that this point is sometimes overlooked, or not fully appreciated when introducing the principles of experimental design. All too often, the exposition jumps quickly, without much further ado, from randomization to the independence and normality assumptions underlying "the linear model" we are all accustomed to be using for analysis (Kempthorne, 1977). One issue that confuses researchers well versed in field experimentation is that the independence assumption seems to be at odds with the real-life experience that close plots are more similar than more distant ones, implying spatial covariance decaying with spatial distance. Just how exactly randomization justifies the independence assumption (Kempthorne, 1952; Mead et al., 2012), despite of this spatial covariance, and how the linear model used for analysis is usually a reasonable approximation, is often not explained in much detail.

### 2.2 Row-column designs

As early as 1924, Fisher was working on Latin squares (Fisher Box, 1978, pp.156-158; Speed 1991), which allow blocking in both rows and columns of a square grid of plots. An empirical comparison, similar to the one reviewed in the previous section, was conducted by Olof Tedin (1931) in very close collaboration with Fisher. According to Tedin, "*The object chiefly held in* *view, besides the elimination of the "systematic" variation in the field, was that different replicates of the same treatment should be removed as far from one another as possible, in order to minimise the soil differences between the different treatments. As the best schemes of arrangement for the 5 × 5 square, the "knight's move" has been devised …*" This layout may be contrasted to the "diagonal arrangement." Even though knight's move Latin squares were expected to have an edge in precision, there was concern regarding the validity of their analysis of variance when the randomization set was restricted to the set of knight's move (or Knut Vik) squares. Tedin investigated this using a criterion he called treatment error coefficient, defined as $t.e.c. = TSS/(TSS + RSS)$, where $TSS$ is the treatment sum of squares and $RSS$ the residual sum of squares. If the ANOVA sums of squares are unbiased, as is the case when the full randomization set is used, the expected value of $t.e.c.$ should be close to the $null\ t.e.c. = TDF/(TDF + RDF)$, where $TDF$ and $RDF$ are the treatment and residual degrees of freedom, respectively. Using uniformity trial data, Tedin found knight's move squares on average to yield $t.e.c.$ below the $null\ t.e.c.$, whereas diagonal squares tend to have $t.e.c.$ above the $null\ t.e.c.$ In fact, as pointed out by Yates (1939), the bias is equal in magnitude and opposite for these two types of Latin square. As a consequence, ANOVA will overestimate the error term for knight's move squares and underestimate it for diagonal squares (Fisher, 1971, §34; Speed, 1991). At the same time, these results confirmed the expectation that knight's move squares yield better precision than diagonal squares.

Work in which we also used $t.e.c.$ started with two papers in 2016 and 2018 (Piepho et al., 2016; Piepho et al., 2018). That work was initiated when Volker Michel, an agronomist at the Landesforschungsanstalt für Landwirtschaft und Fischerei in Mecklenburg-Vorpommern (Germany), contacted me about non-resolvable row-column designs. He routinely uses such designs, preferably with complete blocks in the rows and incomplete blocks in the columns, with the treatment numbers typically ranging between 10 and 30. In his assessment, row and column effects are always expected in the trials they conduct. When using design packages to generate randomized designs, however, he identified two problems that caused him and his colleagues to question the merit of such designs. (i) The same adjacency of treatment pairs often occurred in several rows. This is a problem especially when plots of a variety are prone to loss due to frost kill, in which case the practice at the time was to also discard the two neighbouring plots in the same row due to the expected strong neighbour effects. (ii) Quite frequently the replications of at least one treatment tended to be clustered in one section of the layout, and this did not seem acceptable. Hence, the obvious question was if the randomization could be somehow restricted in order to avoid such undesirable arrangements. We initially devised two different design strategies to achieve what we dubbed neighbour balance (NB) & evenness of distribution (ED), tackling problems (i) and (ii), respectively. One strategy was model-based using specific random-effects models combined with spatial covariance models for the plot error (Piepho et al., 2016), and the other was based on a model-free purely algorithmic approach that excludes certain patterns in a computer-based search (Piepho et al., 2018). The statistical properties of the latter approach were subsequently investigated using Tedin's method based on the $t.e.c.$ (Williams and Piepho, 2018, 2019), confirming approximate validity of the restricted randomization for non-resolvable row-column designs. The algorithmic approach to ensuring NB&ED properties has also been implemented in the CycDesigN package (VSNi, 2025), and it has been extended to other designs (for a review Piepho et al., 2021). This approach may be seen as a late and pragmatic response to a suggestion reportedly made by Fisher in a conversation with L. J. Savage in 1952 (Holschuh, 1980) in relation to Latin squares:

"*What would you do", I had asked, "if, drawing a Latin square at random for an experiment, you happened to draw a Knut Vik square?" Sir Ronald said he thought he would draw again and that, ideally, a theory explicitly excluding regular squares should be developed*. (Savage, 1962, p.88)

Presumably, Fisher's comment primarily had smaller designs in mind, particularly 5 × 5 and 6 × 6 squares (Fisher, 1926). Provided the numbers of plots and treatments are not very small, the randomization set for row-column designs may be so large that the restriction on randomization by systematically excluding certain unwanted configurations may leave enough room for approximately valid inference. As pointed out before, uniformity data provide a great resource for evaluating the validity of any restriction on randomization one is willing to contemplate. A very lucid discussion of settings calling for restrictions on randomization is given in Mead et al. (2012, §11). The pragmatic approach put forward here may be criticized if one insists on the availability of a full randomization theory and strict validity of randomization. This potential criticism should be balanced, however, against the fact that the set of designs with such a full randomization theory is quite limited, and that in practice a much broader set of options is needed.

Piepho et al. (2016) focused on non-resolvable row-column designs, where the full randomization set can contain a substantial number of designs with undesirable NB&ED features. It may be added that resolvable row-column designs usually provide better NB&ED features, especially when combined with latinization, which involves a restriction on randomization in long rows or columns extending across replicates (Piepho et al., 2015). For recent work on latinized row-column designs and the closely related class of semi-Latin squares, see Williams and Bailey (2026).

### 2.3 Augmented row-column designs

In some settings, the material available for the treatments of interest is so limited that no replication is feasible. One common example is early generation plant breeding trials, where seed availability of test lines is so small that only one plot per trial and treatment can be accommodated. Federer (1956, 1961) proposed to employ check varieties (checks) with sufficient seed to provide replication and sufficient information for making block adjustments. The key idea is to use a good blocked design for a limited number of checks and to augment these blocks with unreplicated test lines. This general idea can be applied with various forms of blocking, including row-column designs. Such augmented designs can nowadays be conveniently generated using computer-based search strategies combined with theoretical results, such as upper bounds of the average efficiency factor (John and Williams, 1995, §2) or overall efficiency factor (Bailey, 2008, §11.8), computed from Fisher's information matrix for treatments.

Recently, work has been done focussing on specific configurations as regards the number of rows, columns, number of checks and their replication. Bailey and Haines (2025) consider square layouts for augmented designs and how these can be generated from auxiliary designs, also known as contractions (Patterson and Williams, 1976a). This interesting work can be linked back to a report by Fisher (1938) explaining how a Youden square, which has a rectangular layout with the numbers of rows and columns differing, can be represented as a square array design (Cameron, 2003; Bailey and Haines, 2025, § 3.1). Piepho and Williams (2026) extended the work of Bailey and Haines (2025) to rectangular layouts. Under certain conditions on the dimensions of the layout, an explicit equation can be given for the average efficiency factor of the augmented row-column design that depends solely on the dimensions of the contraction from which it is generated. The generation of augmented designs from a contraction is much quicker than direct generation using computer search, and as shown in two examples in our paper, designs generated from contractions can be quite competitive in terms of the average efficiency factor. From a practical point of view, the availability of designs generated from contractions, with known theoretical properties, provides a very useful benchmark against which general-purpose search strategies providing augmented row-column designs for a wider range of settings may be compared.

## 3. Multi-environment trials

### 3.1 Basics

After having covered various designs for individual experiments and how to analyse these, Fisher (1971, §65) discusses MET, emphasizing that treatment-environment interaction must be considered as the error term for comparing treatment means. It is plainly clear from his writing that he regarded environment to be a random factor in this context. Moreover, he emphasized that the overall precision of a MET depends less on the precision of the individual experiments but mainly on the number of sites, especially if the treatment-by-environment interaction is large relative to the error variance. He made these statements in terms of ANOVA mean squares rather than in terms of variance components of a stated linear model, but clearly he must have implicitly employed a linear mixed model with random effects for interaction to arrive at these statements. This is particularly apparent in the passage where he considers the case of an experiment conducted on ten farms for five years (seasons). In one key paragraph (Fisher, 1971, p.211) he stresses that

> *the categories [...] "farms" and "seasons" may be described as indefinite. By this is meant that although the experiment may have shown that the conditions on the different farms are not equivalent, yet we have no means of reproducing such different conditions at will. If the results of the experiment are used to predict the responses to be expected at other farms, chosen at random from the same population, these will probably differ to much the same extent from those of the original experiment, and the interaction with farms supplies the appropriate estimate of error.*

This view very closely corresponds to the concept of a target population of environments (TPE), a term coined by Comstock and Moll (1963), and still in common usage up to this date. Then turning to the appropriate error term when seasons are taken into account as well, Fisher states the following (Fisher, 1971, p.213):

> *When the significance neither of the differences between farms, nor those between years, is in doubt, a satisfactory estimate of the variance of the general mean is found by adding the mean squares for farms and for years, subtracting that for interaction, and dividing by 50.*

Fisher further hints that "*The error is not exactly of "Student"'s type, so that no number of degrees of freedom should be assigned to it.*" Satterthwaite's method, as well as its generalization known as Kenward-Roger method, needed yet to be conceived.

In relation to the general mean, it is implicitly clear that he must have referred to a single treatment, for which the following mixed model can be assumed:

$$y_{ijk} = \mu + f_i + s_j + (fs)_{ij} + e_{ijk} \quad (1)$$

where $y_{ijk}$ $(i = 1,...,m; j = 1,...,n; k = 1,...,r)$ is the response of the $k$-th replicate on the $i$-th farm (site) in the $j$-th season (year), $\mu$ is the general mean, $f_i \sim N(0, \sigma_f^2)$ is a random main effect for the $i$-th farm, $s_i \sim N(0, \sigma_s^2)$ is a random main effect for the $j$-th season, $(fs)_{ij} \sim N(0, \sigma_{fs}^2)$ is a random farm-by-season interaction effect, and $e_{ijk} \sim N(0, \sigma^2)$ is a residual error term. The ANOVA for data described by this mixed model is shown in Table 1.

**Table 1**: Analysis of variance based on the model in eq. (1)

| Source | Degrees of freedom[§] | Mean square | Expected mean square |
|---|---|---|---|
| | | | |
| General mean | 1 | $MS_{\mu}$ | $\sigma^2 + r\sigma_{fs}^2 + rn\sigma_f^2 + rm\sigma_s^2 + rmn\mu^2$ |
| Farms | $(m-1)$ | $MS_F$ | $\sigma^2 + r\sigma_{fs}^2 + rn\sigma_f^2$ |
| Seasons | $(n-1)$ | $MS_S$ | $\sigma^2 + r\sigma_{fs}^2 + rm\sigma_s^2$ |
| Farms × Seasons | $(m-1)(n-1)$ | $MS_{F \times S}$ | $\sigma^2 + r\sigma_{fs}^2$ |
| Error | $mn(r-1)$ | $MS_R$ | $\sigma^2$ |

§ $m$ = number of farms; $n$ = number of seasons, $r$ = number of replications per farm and season

The variance of the estimator of the general mean, $\hat{\mu} = (mnr)^{-1} \sum_{i=1}^{m} \sum_{j=1}^{n} \sum_{k=1}^{r} y_{ijk}$, where $m$, $n$ and $r$ are the numbers of farms, seasons and replicates, respectively, is given by

$$\text{var}(\hat{\mu}) = \frac{\sigma_f^2}{m} + \frac{\sigma_s^2}{n} + \frac{\sigma_{fs}^2}{mn} + \frac{\sigma^2}{mnr} \quad , \quad (2)$$

which, based on the expected mean squares in Table 1, can be estimated by

$$est.\text{var}(\hat{\mu}) = \frac{MS_F + MS_S - MS_{F \times S}}{mnr} \quad . \quad (3)$$

Comparing this to Fisher's suggestion for the example with ten farms and five seasons, we may speculate as to the exact scenario he had in mind, given that his divisor for the estimated variance is $mn = 50$, implying that there is only one observation to be analysed per farm and season $(r = 1)$. One can think of the following three scenarios, all of which lead to his estimator in eq. (3): (i) There is only a single plot per farm and season. (ii) We analyse means over replicate plots per farm and season. (iii) Analysis is based on sums over replicate plots per farm and season.

The precision considered so far is that for a single general mean, i.e. a single treatment. This is readily extended to experiments with several treatments. Assuming a completely randomized design for simplicity, we may extend the model in eq. (1) by adding a main effect for treatments and all interactions of treatments with farms and seasons:

$$y_{hijk} = \mu + \tau_h + f_i + s_j + (fs)_{ij} + (\tau f)_{hi} + (\tau s)_{hj} + (\tau fs)_{hij} + e_{hijk} \quad (4)$$

where $\tau_h$ is the main effect of the $h$-th treatment $(h = 1,...,t)$, $(\tau f)_{hi} \sim N(0, \sigma^2_{\tau f})$ is the random treatment-by-farm interaction effect, $(\tau s)_{hj} \sim N(0, \sigma^2_{\tau s})$ is the random treatment-by-season interaction effect, $(\tau fs)_{hij} \sim N(0, \sigma^2_{\tau fs})$ is the random treatment-by-farm-by-season interaction effect, and all other terms are as defined for eq. (1). The mean of the $h$-th treatment is given by $\mu_h = \mu + \tau_h$, and half the variance of a difference (the "semi-variance") of the estimated means of two treatments $h$ and $h'$ (*cf.* Talbot, 1984):

$$sv = \frac{\sigma^2_{\tau f}}{m} + \frac{\sigma^2_{\tau s}}{n} + \frac{\sigma^2_{\tau fs}}{mn} + \frac{\sigma^2}{mnr} \quad . \quad (5)$$

**Table 2**: Analysis of variance based on the model in eq. (4). Note that entries for farms, seasons, and farm-by-season interaction have been omitted for brevity (they are the same as those given in Table 1).

| Source | Degrees of freedom§ | Mean square | Expected mean square$ |
|---|---|---|---|
| | | | |
| Treatments | $(t-1)$ | $MS_T$ | $\sigma^2 + r\sigma^2_{\tau fs} + rn\sigma^2_{\tau f} + rm\sigma^2_{\tau s} + rmnQ(\tau)$ |
| Treatments × Farms | $(t-1)(m-1)$ | $MS_{T\times F}$ | $\sigma^2 + r\sigma^2_{\tau fs} + rn\sigma^2_{\tau f}$ |
| Treatments × Seasons | $(t-1)(n-1)$ | $MS_{T\times S}$ | $\sigma^2 + r\sigma^2_{\tau fs} + rm\sigma^2_{\tau s}$ |
| Treatments × Farms × Seasons | $(t-1)(m-1)(n-1)$ | $MS_{T\times F\times S}$ | $\sigma^2 + r\sigma^2_{\tau fs}$ |
| Error | $tmn(r-1)$ | $MS_R$ | $\sigma^2$ |

§ $t$ = number of treatments; $m$ = number of farms; $n$ = number of seasons, $r$ = number of replications per farm and season

\$ $Q(\tau) = (t-1)^{-1}\sum_{h=1}^{t}(\tau_h - \bar{\tau}_\bullet)^2$

Using the three ANOVA mean squares for the interaction of treatments with farms and seasons in Table 2, this semi-variance can be estimated by (*cf.* Moore and Dixon, 2015)

$$est.sv = \frac{MS_{T\times F} + MS_{T\times S} - MS_{T\times F\times S}}{mnr}, \quad (6)$$

which is analogous to eq. (3) and hence to Fisher's proposed estimator for a general mean. It is very likely that Fisher also had eq. (6) in mind when proposing eq. (3) as the estimator of error for the general mean of the response, as all his examples are for comparative experiments. In fact, if we take "the response" to be the pairwise difference among two treatments within a farm and season, eqs. (2) and (3) are equivalent to eqs. (5) and (6), respectively. I have assumed a completely randomized design here for simplicity, but the results given for *sv* and *est.sv* would be the same for individual trials laid out in complete blocks or as Latin squares. An important feature that emerges from eqs. (2) and (5) is that in MET the interaction variances play a dominant role in terms of precision, especially when the within-experiment error variance $\sigma^2$ is small by comparison (Patterson and Silvey, 1980; Speed, 1991).

The semi-variance in eq. (5) does not depend on variance components for pure environmental effects for farms and seasons, and likewise the ANOVA estimator of the semi-variance in eq. (6) does not depend on the corresponding mean squares. Hence, it makes no difference whether these purely environmental effects (farms, seasons, farms-by-seasons interaction) are modelled as fixed or random, even though the status of factors farm and season is clearly random. The main reason why this is the case is that effects operate in an additive fashion according to linear models such as those in eqs. (1) and (4), and that the data is assumed to be balanced. The situation will be different when the set of treatments tested is not the same in all environments as is often the case in variety testing (Patterson and Silvey, 1980). In this case, modelling environmental effects for farms and seasons as random will entail recovery of inter-environment information, in much the same way as random effects for incomplete blocks will allow the recovery of inter-block information in incomplete block designs (Yates, 1940). Incidentally, recovery of information is another seminal idea attributable to Fisher, see Terry Speed's commentary in Samuels et al. (1991). The most important consequence of modelling environmental factors as random is that the interactions with treatment are modelled as random, because these effects then act as the major error terms as evident from eq. (5) and as so clearly spelled out in Fisher (1971, §65). It is mentioned in passing that these same principles are also of importance in arm-based network meta-analysis (Senn, 2023; Piepho et al., 2024).

Yates (1939) stressed the great importance of expressions like that in eq. (5) for optimizing a trialling network. At the same time he makes the interesting point that a valid estimate of error for treatment comparisons may be available with systematic designs under a linear mixed model of the form as in eq. (4), essentially because the error term only depends on treatment-by-environment interaction mean squares:

> *Thus by sacrificing the estimate of error we may succeed in increasing the accuracy of the final means (over a series of years) of the varietal differences at any one place, but not to the same extent as the increase in the accuracy of the individual experiments. At the same time we lose all possibility of effectively estimating the magnitude of the variation due to weather conditions, changes of field, etc. Although this does not invalidate tests of significance of the mean varietal difference (for which the effective error is estimated from varieties × years and varieties × places × years) it is a drawback which may become important as soon as the finer points of varietal improvement begin to be considered.*

It may be surmised that Yates' assessment assumes that, even though systematic designs are assumed to be used in each environment, the treatment labels are permuted in each environment (Piepho et al., 2026). This is in contrast to another practice involving systematic arrangements that I have occasionally come across. In MET for variety assessments, it is sometimes found that the exact same treatment order is used in the first replicate in each environment. One motivation for this practice is facilitation of presentations of trials to visiting audiences. It seems difficult to convince experimenters to deviate from this practice. The most promising route may be to assess the empirical pairwise standard errors, e.g. using uniformity trial data, and demonstrate how these increase with the pairwise distance among treatments in the first replicate and how this, in turn, adversely affects the probability of identifying the best treatments. This latter argument may be particularly helpful where experimenters are adamant that they do not require significance tests or valid standard errors, but are only interested in the best point estimates for making recommendations or selections.

Fisher (1971, §64) also considered the estimation of the variance due to treatments, under "*the wider hypothesis that the yields produced by the different samples are a sample from a normal distribution*." In his book, he did not explicitly consider applications in which the random treatment factor was variety or genotype, even though such applications were surely very close to him, given his central role in the development of quantitative genetics theory, starting with his ground-breaking paper on Mendelian inheritance and the infinitesimal model for complex traits (Fisher, 1918). Such applications are still so central to plant breeding and variety testing today, that it is worth making a link to broad-sense heritability $H^2$, which for a series of experiments carried out at *m* sites in *n* years using individual trials with *r* replicates can be written as (Schmidt et al., 2019)

$$H^2 = \frac{\sigma_\tau^2}{\sigma_\tau^2 + sv}, \tag{7}$$

where $\sigma_\tau^2$ is the variance of treatment effects $\tau_h$, assumed to be identically and independently distributed, and *sv* is the semi-variance given in eq. (5). Whereas agronomists are accustomed to assess precision based on *sv* and related quantities such as the standard error of a difference (s.e.d.), most breeders will grade precision based on $H^2$. It is reassuring to statisticians working on optimal design for MET that the two quantities are so intimately related.

Fisher and Mackenzie (1923), in the second half of their seminal paper on analysis of variance, suggested that treatment effects in factorial designs may be better described by a multiplicative model than by an additive model (also see Cochran, 1980). They showed how to use eigenvectors and eigenvalues to obtain the least-squares fit for a model of the form $y_{hij} = \alpha_h \beta_i + e_{hij}$, where $\eta_{hi} = \alpha_h \beta_i$ is the expected value of the *h*-th level of the one treatment factor in combination with the *i*-th level of the other treatment factor and $\alpha_h$ and $\beta_i$ are the corresponding effects (Fisher Box, 1978, p.112). Such models and extensions were later used to model genotype-environment interaction in MET. Fisher and Mackenzie clearly had a fixed-effects model in mind for their factorial experiment with varieties of potato and different manures as the two treatment factors, and so did and do many authors using such models for genotype-environment interaction in MET with respect to the factors genotype and environment (e.g. Yates and Cochran, 1938; Kempton, 1984; Gauch, 1988). Seeing such models applied in MET, however, Fisher probably would have argued that the environmental terms should be taken as random if inferences on treatment effects are to be made for a broader TPE. In a MET context this was first done in Gogel et al. (1995), and such models can be cast in terms of factor-analytic variance-covariance structures for genotype-environment interaction, with factor loadings assigned to genotypes, playing the role of sensitivities to latent underlying environmental drivers (Piepho and Williams, 2024). It is with such multiplicative models that the status of factors as fixed or random is particularly crucial. Specifically, if the latent loadings of a factor-analytic variance-covariance structure for genotype-by-environment effects are assigned to the environments rather than to genotypes, as has become quite popular, then environment effectively becomes a fixed factor in this structure and all inferences derived from the MET are necessarily restricted to the specific set of environments in which the individual trials were conducted. I believe many a researcher using such models will benefit from re-reading Fisher (1971, §65) on the importance of modelling environment as a random factor.

### 3.2 Partially replicated designs

Federer's invention of augmented designs (Section 2.3) is a great extension of Fisherian ideas. A further step along these lines was the proposal of partially replicated (p-rep) designs by Cullis et al. (2006). The main motivation for this development was that in augmented designs a large fraction of plots is devoted to checks which may not in themselves be of intrinsic interest to the breeder. If the amount of seed available for the test lines allows some level of replication, then some of the test lines can be duplicated, admitting an estimate of error. Hence, the checks can be replaced with duplicates of test lines, thus focussing all plots on the treatments of interest. If a p-rep design is used in MET, then a different subset of test lines can be duplicated in each individual trial. For example, with 1,000 test lines and five environments, a different subset of 200 test lines can be duplicated in each trial, meaning that each test line will be tested on a total of six plots across the MET. This great idea caught on well with many breeders, and p-rep designs are now widely used, primarily in early generations of public and private breeding programs.

Federer's proposal very closely followed Fisher's principles of blocking and randomization, and over time several variations of augmented designs were proposed that make use of different forms of blocking. When the number of test lines is substantial, as is almost invariably the case in large breeding programs, then the use of incomplete blocks virtually becomes a necessity. Transitioning to p-rep designs, an option is to use an incomplete block design for the duplicates, filling up the blocks with unreplicated test lines. This initial idea can be developed further in the context of MET, aiming to optimize the overall design with respect to test line means over environments. This idea was pursued in Williams et al. (2011), where we made use of α-arrays to generate p-rep designs for MET. The idea of using α-arrays goes back to work on fully replicated designs published exactly half a century ago at the time of this lecture (Patterson and Williams, 1976b). So-called α-designs, a form of resolvable incomplete block design, and various extensions thereof, are now widely used, primarily in variety testing and plant breeding. They greatly expanded the flexibility of incomplete block designs compared to square lattice designs introduced by Yates (1936) and rectangular lattice designs (Harshbarger, 1949). The approach to using α-arrays for the generation of efficient p-rep designs was extended further in Williams et al. (2014) to allow for row-column blocking and providing great flexibility regarding the numbers of treatments, the level of partial replication and the block dimensions. These approaches have been implemented in the CycDesigN package (VSNi, 2025), also providing options for inclusion of check varieties with higher level of replication, and imposing the NB&ED features described in Section 2.2.

Optimizing a p-rep design for MET requires a linear model. To illustrate, consider the case where a resolvable incomplete block design is used for the duplicates at each site in a single year. Here, the linear model can be stated as

$$y_{hijk} = \mu + \tau_h + f_i + (\tau f)_{hi} + r_{ij} + b_{ijk} + e_{hijk} \qquad (8)$$

where effects $\mu$, $\tau_h$, $f_i$, and $(\tau f)_{hi}$ are define as in eq. (4), $r_{ij}$ is the effect of the $j$-th replicate at the $i$-th site, and $b_{ijk}$ is the effect of the $k$-th incomplete block in the $j$-th replicate at the $i$-th site. Following Fisher, site may be considered as a random factor, in which case all effects indexed by site would be modelled as random as well, the important consequence being that the random interaction $(\tau f)_{hi}$ becomes the primary error term for comparing treatment means. Given that this effect dominates the precision of treatment comparisons, one may consider that the optimal allocation of treatments to incomplete blocks matters most when the variance of the interaction $(\tau f)_{hi}$ is zero. From a design perspective, this setting can therefore be regarded as the worst-case scenario. This prompted us to use the reduced model

$$y_{hijk} = \mu + \tau_h + f_i + r_{ij} + b_{ijk} + e_{hijk} \quad (9)$$

for optimizing the average efficiency factor of p-rep designs for MET (Williams et al., 2011). In keeping with standard practice for the generation incomplete block designs, all effects except the residual plot error $e_{hijk}$ in eq. (9) were considered as fixed. The great advantage of this approach is that no prior values need to be assigned to the variance components. The fact that the inter-site information is not utilized in this approach can well be tolerated, because the variance for the site main effect $f_i$ is usually large, meaning that there is little information to be recovered. Moreover, in p-rep designs all test lines are tested at each site, also implying that the inter-site information is usually minimal.

### 3.3 Optimal allocation for a sub-divided target population of environments

An important extension of Fisher's view of sites and years in a MET as representing a sample of environments from a larger population was perhaps first put forward by Comstock and Moll (1963), who suggested that a TPE may be stratified into several sub-regions or zones, such that genotype-environment interaction is relatively small within each of the zones. Potentially there can be different connected breeding programs or sub-programs targeting each zone. Similarly, in variety trials, recommendations may be focused on specific agro-ecological zones within a larger TPE. If the same varieties or breeding lines are evaluated in each of the zones, the option of borrowing information between zones becomes relevant. This idea was developed by Atlin et al. (2000), who invoked the concepts of broad-sense heritability [see Section 3.1, eq. (7)] and of correlated response to selection (Falconer, 1952; Curnow, 1988) in order to assess the value of correlated information borrowed between zones. The idea of genetic correlation between different environments was popularized by Falconer (1952). His idea, in turn, bears some resemblance to Fisher's seminal work on genetic correlation among relatives (Fisher, 1918), and can be regarded as a direct extension thereof (Reeves, 1952).

In order to exploit correlation among zones in terms of the performance of the same treatment in different zones, it is necessary to model treatment as a random factor. For illustration, consider the case of a single year's data and a TPE subdivided into zones. Assuming the design in the individual trials is an RCBD, the linear model can be stated as

$$y_{hi(p)jk} = \mu_p + \tau_{h(p)} + f_{i(p)} + (\tau f)_{hi(p)} + r_{i(p)j} + e_{hi(p)jk} \quad (10)$$

where all effects, defined analogously to those in eq. (8), carry an index $p$ $(p = 1,...,q)$ for the zone, in other words all effects are nested within zones. Collecting the zone-specific random effects of the $h$-th genotype into a vector $\boldsymbol{\tau_h} = (\tau_{h(1)}, \tau_{h(2)},..., \tau_{h(q)})^T$, different variance-covariance structures may be imposed for the variance-covariance matrix $\text{var}(\boldsymbol{\tau_h})$, including compound symmetry, factor-analytic or unstructured. Using best linear unbiased prediction (BLUP) of $\boldsymbol{\tau_h}$, the correlation between zones is fully exploited and information can be borrowed from other zones when making predictions for a specific zone (Piepho and Möhring, 2005).

An important design problem in this context is the optimal allocation of trials to zones, given a fixed budget for the total number of trials that can be accommodated. The problem has similarities to that in stratified sampling for censuses and surveys (Yates, 1981). Using the mean squared error of BLUP for treatment effects or pairwise contrasts thereof as optimality criterion, it can be shown that if a simple variance-components model is assumed for all random effects, and a compound symmetry structure for $\mathrm{var}(\boldsymbol{\tau}_h)$, the optimal allocation is the equal split between zones (Prus and Piepho, 2021). With other structures, the optimal allocation differs from this simple design, and the optimum depends on the values of variances and covariances in $\mathrm{var}(\boldsymbol{\tau}_h)$. Intuitively, one may expect that for two highly correlated zones a smaller sample size is needed per zone than for a zone that is not closely correlated with any other zone in the TPE. This is exactly what we find using the analytical approach detailed in Prus and Piepho (2021). Various extensions have been considered. Prus and Piepho (2026) extended the approach to multi-year data. Prus (2025) and Bodner and Prus (2026) showed how the optimal allocation can be tackled when the test lines are correlated due to pedigree or kinship. Further extensions of the general approach are underway, using a combination of analytical results where available, and numerical optimization as well as computer search where necessary.

### 3.4 Merging the trialling systems of two countries

Many European countries have statutory trialling systems to assess the value for cultivation and use (VCU) of new crop varieties. The design for these VCU trials is quite comparable between countries, and analysis is routinely done using linear mixed models like those reviewed in Section 3.1 (Ramakers et al., 2026). The individual trial designs are fully replicated, with RCBDs still being used in most cases and α-designs (Patterson and Williams, 1976b) in some instances. Even though released varieties are often suitable across several countries in Europe, breeders typically submit variety candidates only to one national examination office. Once the variety is added to the National List in one country, it can be marketed in other European countries as well.

This current system means that there is very little overlap between countries in the set of varieties tested in any given year. This arrangement is sub-optimal when it comes to making recommendations to farmers, especially in countries where a variety was not tested in VCU trials. In such countries, the evaluation in official variety testing needs to essentially start from scratch. Given that solid recommendations can be made only after several years of testing [see eq. (5)], transfer of breeding progress to practical farming is slower than it would be with VCU testing extended to all countries in which a variety candidate is deemed adapted.

Optimizing the overall design when integrating the trialling systems of several countries poses a number of challenges, and finding the statistical optimum is but one of these challenges. Given that no fully developed methods are as yet available for this complex design problem, progress can be made by empirically exploring the relative merits of different design options that suggest themselves. A very powerful framework for such evaluations was proposed by Stroup (2002). The key idea is that a mixed model package can be used to conveniently compute the non-centrality parameter of an F-distribution for significance tests of interest, from which the power can be readily computed. What is required is a dataframe that represents the contemplated design, and response values designed such that they represent the values of the fixed effects of interest for a pre-specified effect size. Furthermore, variance parameter values must be supplied, which are entered into the package via options for starting value specifications for residual maximum likelihood estimation (REML) (Patterson and Thompson, 1971). If the package allows suppressing REML iterations, and hence fixing variance parameters at their start values, the non-centrality parameter can be directly computed for the assumed design and values of the variance parameters. The focus in Stroup (2002) is indeed on the power of F-tests, which links it very closely to Fisher's seminal contributions around analysis of variance and the F-distribution. The approach can also be used for the more modest objective of assessing the precision (or Fisher information) of estimators of effects of interest. In light of the great versatility of this approach, it is surprising that it does not yet seem to have gained widespread usage in practice.

Piepho and Malik (2025) considered the trialling systems of Germany and Poland for maize. Using variance component estimates obtained in a joint analysis of multi-year VCU trial data from both countries, several design options were compared using the approach of Stroup (2002). There was some overlap between countries of varieties over the years, allowing genetic correlations to be estimated using models similar to the one given in Section 3.3, regarding the two countries as two different zones. Separate analysis for the individual countries, which may be regarded as the baseline, can be assessed setting the genetic correlation equal to zero. A joint analysis, exploiting the genetic correlation among countries via BLUP, already provides major gains in precision, even without any alteration of the MET design. Further gains are possible by changes to the design. The major change of design is to test all varieties in both countries. Leaving the total capacity of the trialling systems of both countries unchanged, this means reducing the intensity of testing per variety-country combination. The reduction is more than offset by the gain arising from the ability to borrow information between countries, especially when genetic correlation is high. The most extreme scenario in terms of departure from the current system considered in Piepho and Malik (2025) tests each variety in each individual trial, which requires using p-rep designs as discussed in Section 3.2. This design was the most efficient overall among the scenarios considered. It cannot, of course, be regarded as the optimal design, but in lack of an established framework that would directly return the optimal design, the trial-and-error approach using Stroup's method provides a good start towards improvement of the current system.

## 4 Conclusions

Alluding to Fisher's foundational work on the design of experiments in several places, his three famous principles in particular, this lecture has considered several issues around the design of field trials, starting from individual trials in Section 2 and transitioning to MET in Section 3. The examples given towards the end highlight the fact that design problems can be particularly multi-faceted when it comes to MET. There is certainly an urgent need for further research on optimal design especially for MET. In closing, only a few topics for future research are briefly mentioned.

Plant breeding programs require optimization with respect to several parameters. Only some of these are amenable to optimization based on classical approaches for the design of experiments. Apart from the choice of individual trial designs and the number of trials to be conducted in each zone of interest, breeders need to take decisions around the particular set of genotypes to include in the trials in any one generation of the full breeding cycle, and how many lines to select for further testing and evaluation in the next year in each generation (Hoefler et al., 2020). A further related question concerns the number of check genotypes to include in order to connect the data across multiple generations or breeding cycles (Piepho and Laidig, 2025). Moreover, there is an increasing interest in approaches that allow incorporating environmental covariate information, readily available in public databases that draw on various sources (satellites, remote sensing, drones, etc.), in selection decisions (Resende et al., 2022), and implications of the incorporation of such information for the design of MET are as yet entirely unexplored. Here, cross-fertilization from other areas of experimental design where the inclusion of covariates is customary, such as computer experiments (Santner et al., 2003), industrial experiments (Goos and Jones, 2011) or clinical trials (Rosenberger, 2026) will definitely be beneficial. Almost invariably, the overall optimum for a breeding programme will depend, among other things, on genetic relationships among the lines in the breeding population under test. Solving these complex optimization problems in practice will require new methodological developments (Heslot and Feoktistov, 2020).

Even though the ultimate goal ideally is the identification of an overall optimal design, lacking a fully-fledged framework for locating the optimum, good progress can often be made based on some educated guesses, drawing on Fisher's principles, as to the most promising way forward (Mead et al., 2012). Approaches like that proposed by Stroup (2002) can be a great aid in evaluating competing options. A good current example is what has been dubbed sparse testing, a term that refers to MET designs, where each genotype is only tested in a subset of environments (Crespo-Herrera et al., 2021). Sparse testing may be viewed as an extension of the p-rep idea, where testing intensity is thinned out so much that the average number of replicates per treatment and trial falls below unity. A key aspect of sparse testing is that information can be borrowed across test lines via the pedigree or kinship via best linear unbiased prediction (Henderson, 1984; Meuwissen et al., 2001), which is rooted in Fisher (1918), or a host of contending methods, including Bayesian prediction models (Pérez and de los Campos, 2014) and machine learning methods (Montesinos-López et al., 2024). The efficiency of all of these prediction methods depends crucially on the design of the MET in terms of the choice of genotypes and their allocation to individual trials, and on how the design is accounted for in the analysis. In a simple breeding population with multiple families and multiple test lines per family, e.g., good efficiency is feasible so long as each family is represented with a few test lines in each trial. The first proposals for sparse testing connected trials by replicating a small select subset of test lines in all environments and testing the remaining lines in just a small subset of the environments. Initially, little thought was given to the question of how exactly test lines should be allocated to environments in an optimal way. Taking recourse to the idea of efficient blocking, and regarding environments as incomplete blocks, much progress can be made in sparse testing (Montesinos-López et al., 2022). Further enhancement should be possible by explicitly taking the pedigree into account using a model with random treatment effects correlated according to familial relationships (Fisher, 1918; Bueno-Filho and Gilmour, 2007). Ultimately, more rigorous options to overall design optimization are certainly desirable.

## Acknowledgements

Hans-Peter Piepho was selected as the 43rd Memorial Lecturer by the Fisher Memorial Trust. He sincerely thanks the Fisher Memorial Trust for this invitation, which is truly a great honour. This paper was read at King's College London on April 23, 2026. The audience included R. A. Fisher's granddaughter, the Rev. Jenny Tebboth, and doctoral student Sir Walter Bodmer.

*Conflict of interest*: None declared.

## Data availability

There are no data in this paper.

## References

Alesso C. A., Cipriotti P. A., Bollero G. A., & Martin N. F. (2021). Design of on-farm experiments to estimate site-specific crop responses. *Agronomy Journal*, *113*, 1366–1380. https://doi.org/10.1002/agj2.20572

Atlin G. N., Baker R. J., McRae K. B. & Lu X. (2000). Selection response in subdivided target regions. *Crop Science*, *40*, 7–13. https://doi.org/10.2135/cropsci2000.4017

Bailey R. A. (2008). *Design of comparative experiments*. Cambridge University Press.

Bailey R. A., & Haines L. M. (2025). Square array designs for unreplicated test treatments with replicated controls. *Journal of Agricultural Biological and Environmental Statistics*. https://doi.org/10.1007/s13253-025-00699-1

Bodmer W., Bailey R. A., Charlesworth B., Eyre-Walker A., Farewell V., Mead A., & Senn S. (2021). The outstanding scientist, R. A. Fisher: his views on eugenics and race. *Heredity*, *126*, 565–576.

Bodner T., & Prus M. (2026). Dimension reduction for optimal design problems with Kronecker product structure. *Scandinavian Journal of Statistics*. https://doi.org/10.1111/sjos.70061

Bueno Filho J. S. D. S., & Gilmour S. G. (2007). Block designs for random treatment effects. *Journal of Statistical Planning and Inference*, *137*, 1446–1451.

Cameron P. (2003). Multi-letter Youden rectangles from quadratic forms. *Discrete Mathematics*, *266*, 143–151.

Cao Z., Brown J., Gibberd M., Easton J., & Rakshit S. (2024). Optimal design for on-farm strip trials—systematic or randomised? *Field Crops Research*, *318*, 109594. https://doi.org/10.1016/j.fcr.2024.109594

Cochran W. G. (1980). Fisher and the analysis of variance. p.17–34. In: Fienberg S. E., & Hinkley D. V. (eds.) *R. A. Fisher*: *An appreciation*. Springer.

Comstock R. E., & Moll R. H. (1963). Genotype-environment interaction. p. 164–194. *In* Hanson W. D., & Robinson H. F. (eds.) Statistical genetics and plant breeding. Publication 982. National Academy of Sciences - National Research Council, Washington, DC.

Cox D. R., & McCullagh P. (1982). Some aspects of analysis of covariance. *Biometrics*, *38*, 541–561.

Crespo-Herrera L., Howard R., Piepho H. P., Pérez-Rodríguez P., Montesinos-Lopez O., Burgueño J., Singh R., Mondal S., Jarquin D., & Crossa J. (2021). Genomic-enabled prediction for sparse testing in wheat multi-environmental trial. *The Plant Genome*, *14*, e20151. https://doi.org/10.1002/tpg2.20151

Cressie N. A. C., & Wikle C. K. (2011). *Statistics for spatio-temporal data*. Wiley.

Cullis B. R., Smith A. B., & Coombes N. E. (2006). On the design of early generation variety trials with correlated data. *Journal of Agricultural, Biological and Environmental Statistics*, *11*, 381–393. https://doi.org/10.1198/108571106X154443

Curnow R. N. (1988). The use of correlated information on treatment effects when selecting the best treatment. *Biometrika*, *75*, 287–293. https://doi.org/10.1093/biomet/75.2.287

De Faveri J., Verbyla A. P., Cullis B. R., Pitchford W. S., & Thompson R. (2017). Residual variance covariance modelling in analysis of multivariate data from variety selection trials. *Journal of Agricultural, Biological & Environmental Statistics*, *22,* 1–22. https://doi: 10.1007/s13253-016-0267-0
Falconer D. S. (1952). The problem of environment and selection. *American Naturalist*, *86*, 293–298. https://doi.org/10.1086/281736
Federer W. T. (1956). Augmented (or hoonuiaku) designs. *Hawaiian Sugar Planters Record*, *55*, 191–208.
Federer W. T. (1961). Augmented designs with one-way elimination of heterogeneity. *Biometrics*, *17*, 447–473.
Fisher R. A. (1918). The correlation between relatives on the supposition of Mendelian inheritance. *Transactions of the Royal Society Edinburgh, 52*, 399–433. https://doi.org/10.1017/S0080456800012163
Fisher R. A. (1925). Statistical methods for research workers. Oliver & Boyd.
Fisher R. A. (1926). The arrangement of field experiments. *Journal of the Ministry of Agriculture of Great Britain*, *33*, 503−513.
Fisher R. A. (1935). *Design of experiments*. Oliver & Boyd.
Fisher R. A. (1938). The mathematics of experimentation. *Nature*, *142*, 442−443.
Fisher R. A. (1971). *Design of experiments*. Eighth edition. Reprinted by arrangement. Hafner Publishing Company.
Fisher R. A. (1973). *Statistical methods for research workers*. Fourteenth edition – revised and enlarged. Hafner Publishing Company.
Fisher R. A., & Mackenzie, M. A. (1923). Studies in crop variation. II. The manorial response of different potato varieties. *Journal of Agricultural Science*, *13*, 311−320. https://doi.org/10.1017/S0021859600003592
Fisher Box J. (1978). *R. A. Fisher. The life of a scientist*. Wiley.
Gauch H. G. (1988). Model selection and validation for yield trials with interaction. *Biometrics*, *44*, 705−715. https://doi.org/10.2307/2531585
Gogel B., Cullis B. R., & Verbyla A. P. (1995). REML estimation of multiplicative effects in multi-environment variety trials. *Biometrics*, *51*, 744−749. https://doi.org/10.2307/2532960
Goos P., & Jones B. (2011). *Optimal design of experiments*. Wiley.
GRDC – Grains Research and Development Cooperation (2021). GRDC Precision Agriculture Manual. Designing your own on-farm experiments. https://grdc.com.au/resources-and-publications/all-publications/publications/2021/designing-your-own-on-farm-experiments
Harshbarger B. (1949). Triple rectangular lattices. *Biometrics*, *5*, 1−15. https://doi.org/10.2307/3001888
Hartung J. et al. (2026). (in preparation)
Henderson C. R. (1984). *Applications of linear models in animal breeding*. University of Guelph.
Heslot N., & Feoktistov V. (2020). Optimization of selective phenotyping and population design for genomic prediction. *Journal of Agricultural, Biological, and Environmental Statistics*, *25*, 579−600. https://doi.org/10.1007/s13253-020-00415-1
Hoefler R., González-Barrios P., Bhatta M., Nunes J. A. R., Berro I., Nalin R. S., Borges A., Covarrubias E., Diaz-Garcia L., Quincke M., & Gutierrez L. (2020). Do spatial designs outperform classic experimental designs? *Journal of Agricultural, Biological, and Environmental Statistics*, *25*, 523−552. https://doi.org/10.1007/s13253-020-00406-2
Holschuh N. (1980). Randomization and design: I. p.35−45. In: Fienberg S. E., & Hinkley D. V. (eds.) *R. A. Fisher*: *An appreciation*. Springer.
John J. A., & Williams E. R. (1995). *Cyclic and computer generated designs* (2nd ed.). Chapman and Hall.
Kempthorne O. (1952). *The design and analysis of experiments*. Wiley.

Kempthorne O. (1977). Why randomize. *Journal of Statistical Planning and Inference*, *1*, 1–25. https://doi.org/10.1016/0378-3758(77)90002-7
Kempton R. A. (1984). The use of biplots in interpreting variety by environment interactions. *Journal of Agricultural Science*, *103*, 123–135. https://doi.org/10.1017/S0021859600043392
Macholdt J., Piepho H. P., & Honermeier B. (2019a). Long-term impact of sub-optimal and optimal nutrient supply on grain yield and yield stability of winter wheat. *European Journal of Agronomy*, *102*, 14–22. https://doi.org/10.1016/j.eja.2018.10.007
Macholdt J., Piepho H.P., & Honermeier B. (2019b). Does fertilization impact production risk and yield stability across an entire crop rotation? Insights from a long-term experiment. *Field Crops Research*, *238*, 82–92. https://doi.org/10.1016/j.fcr.2019.04.014
Mead R., Gilmour S. G., & Mead A. (2012). *Statistical principles for the design of experiments. Applications to real experiments*. Cambridge University Press.
Mercer W. B., & Hall A. D. (1911). The experimental error of field trials. *Journal of Agricultural Science*, *4*, 107–132. https://doi.org/10.1017/S002185960000160X
Meuwissen T. H., Hayes B. J., Goddard M. (2001). Prediction of total genetic value using genome-wide dense marker maps. *Genetics*, *157*, 1819–29.
Mitscherlich E. A. (1919). *Vorschriften zur Anstellung von Feldversuchen in der landwirtschaftlichen Praxis*. Berlin: Verlag Paul Parey.
Mitscherlich E. A. (1925). *Vorschriften zur Anstellung von Feldversuchen in der landwirtschaftlichen Praxis*, 2. Auflage. Berlin: Verlag Paul Parey.
Montesinos-López O. A., Chavira-Flores M., Kismiantini , Crespo-Herrera L., Saint Piere C., Li H., Fritsche-Neto R., Al-Nowibet K., Montesinos-López A., Crossa J. (2024). A review of multimodal deep learning methods for genomic-enabled prediction in plant breeding. *Genetics*, *228*, iyae161. https://doi.org/10.1093/genetics/iyae161
Montesinos-López O. A., Montesinos-López A., Acosta R., Varshney R. K., Bentley A., Crossa J. (2022). Using an incomplete block design to allocate lines to environments improves sparse genome-based prediction in plant breeding. *Plant Genome*, *15*, e20194. https://doi.org/10.1002/tpg2.20194
Moore K. J., & Dixon P. (2015). Analysis of combined experiments revisited. *Agronomy Journal*, 107, 763–771. https://doi.org/10.2134/agronj13.0485
Patterson H. D., & Silvey V. (1980). Statutory and recommended list trials of crop varieties in the United Kondom. *Journal of the Royal Statistical Society A*, *143*, 219–252. https://doi.org/10.2307/2982128
Patterson H. D., & Thompson R. (1971). Recovery of inter-block information when block sizes are unequal. *Biometrika*, *58*, 545–554. https://doi.org/10.1093/biomet/58.3.545
Patterson H. D., & Williams E. R. (1976a). Some theoretical results on general block designs. *Proceedings of the 5th British Combinatorial Conference. Congressus Numerantium*, *XV*, 489–496.
Patterson H. D., & Williams E. R. (1976b). A new class of resolvable incomplete block designs. *Biometrika*, *63*, 83–92. https://doi.org/10.1093/biomet/63.1.83
Pérez P., & de los Campos G. (2014). Genome-wide regression and prediction with the BGLR statistical package. *Genetics*, *198*, 483–495. https://doi.org/10.1534/genetics.114.164442
Picard R. (1980). Randomization and design: II. p.46–58. In: Fienberg S. E., & Hinkley D. V. (eds.) *R. A. Fisher*: *An appreciation*. Springer.
Piepho H. P., & Laidig F. (2025). How many checks are needed per cycle in a plant breeding or variety testing program? *Plant Breeding*, *144*, 242–248. http://doi.org/10.1111/pbr.13240
Piepho H. P., Madden L. V., & Williams E. R. (2024). The use of fixed study main effects in arm-based network meta-analysis. *Research Synthesis Methods*, *15*, 747–750. https://doi.org/10.1002/jrsm.1721

Piepho H. P., & Malik W. A. (2025). Connecting variety trialling systems across countries. *Plant Breeding*, *144*, 257–262. https://doi.org/10.1111/pbr.13243
Piepho H. P., Malik W. A., & Williams E. R. (2026). An empirical comparison of systematic and randomized field experimental designs. *Journal of Agricultural Science, 164*, e36. https://doi.org/10.1017/S0021859626100732
Piepho H. P., Michel V., & Williams E. R. (2015). Beyond Latin squares: A brief tour of row-column designs. *Agronomy Journal*, *107*, 2263–2270. https://doi.org/10.2134/agronj15.0144
Piepho H. P., Michel V., & Williams E. R. (2018). Neighbour balance and evenness of distribution of treatment replications in row-column designs. *Biometrical Journal*, *60*, 1172–1189. https://doi.org/10.1002/bimj.201800013
Piepho H. P., & Möhring J. (2005). Best linear unbiased prediction for subdivided target regions. *Crop Science*, *45*, 1151–1159. https://doi.org/10.2135/cropsci2004.0398
Piepho H. P., Richter C., Spilke J., Hartung K., Kunick A., & Thöle H. (2011). Statistical aspects of on-farm experimentation. *Crop and Pasture Science*, *62*, 721–735. https://doi.org/10.1071/CP11175
Piepho H. P., & Vo-Thanh N. (2020). Die Gleitmethode nach Mitscherlich und was sie mit Geostatistik zu tun hat. *Journal für Kulturpflanzen*, *72*, 527–540. https://doi.org/10.5073/JfK.2020.10-11.03
Piepho H. P., & Williams E. R. (2024). Factor-analytic variance-covariance structures for prediction into a target population of environments. *Biometrical Journal*, *66*, e202400008. https://doi.org/10.1002/bimj.202400008
Piepho H. P., & Williams E. R. (2026). Rectangular augmented row-column designs generated from contractions. *Biometrical Journal* (accepted; please add DOI)
Piepho H. P., Williams E. R., & Michel V. (2016). Nonresolvable row-column designs with an even distribution of treatment replications. *Journal of Agricultural, Biological and Environmental Statistics*, *21*, 227–242. https://doi.org/10.1007/s13253-015-0241-2
Piepho H. P., Williams E. R., & Michel V. (2021). Generating row-column field experimental designs with good neighbour balance and even distribution of treatment replications. *Journal of Agronomy and Crop Science*, *207*, 745–753. https://doi.org/10.1111/jac.12463
Prus M. (2025). Computing optimal allocation of trials to subregions in crop-variety testing in case of correlated genotype effects. *Statistica Neerlandica*, *79*, e12353. https://doi.org/10.1111/stan.12353
Prus M., & Piepho H. P. (2021). Optimizing the allocation of trials to sub-regions in multi-environment crop variety testing. *Journal of Agricultural Biological and Environmental Statistics*, *26*, 267–288. https://doi.org/10.1007/s13253-024-00659-1
Prus M., & Piepho H. P. (2026). Optimizing the allocation of trials to sub-regions in multi-environment crop variety testing for multi-year experiments. *Journal of Agricultural Biological and Environmental Statistics*, *31*, 207–307. https://doi.org/10.1007/s13253-024-00659-1
Ramakers J. J. C., Malik W. A., Welcker C., Parent B., Bustos-Korts D., Abu-Samra Spencer N., Buntaran H., Chen X., Eylenbosch D., Flamm C., Grizeau C., Herrera J. M., Hiltbrunner J., Horáková V., Levy Häner L., Masson F., Pannecoucque J., Pécs M., Povolný M., Starnberger P., Visse-Mansiaux M., Vonlanthen T., Collonnier C., Laurens F., Martre P., Piepho H. P., & van Eeuwijk F. A. (2026). Evaluation and optimization of wheat and maize national evaluation systems in Europe. *Field Crops Research*, 341, 110420. https://doi.org/10.1016/j.fcr.2026.110420
Reeves E. C. R. (1952). Expression of genes affecting a quantitative character in two different environments. *Heredity*, *6*, 280. https://doi.org/10.1038/hdy.1952.39

Resende R. T., Chenu K., Rasmussen S. K., Heinemann A. B., & Fritsche-Neto R. (2022). Editorial: Enviromics in plant breeding. *Frontiers in Plant Science*, *13*, 935380. https://doi.org/10.3389/fpls.2022.935380

Richter C., & Kroschweski B. (2012). Geostatistical models in agricultural field experiments: Investigations based on uniformity trials. *Agronomy Journal*, *104*, 91–105. https://doi.org/10.2134/agronj2011.0100

Rosenberger W. F. (2026). The 41st Fisher Memorial Lecture. From Fisher to CARA: the evolution of randomization and randomization-based inference. *Journal of the Royal Statistical Society A*, *189*, 497–511. https://doi.org/10.1093/jrsssa/qnaf002

Samuels M. L., Casella G., & McCabe G. P. (1991). Interpreting blocks as random factors. *Journal of the American Statistical Association*, *86*, 798–821. https://doi.org/10.2307/2290415

Santner T. J., Williams B. J., & Notz W. I. (2003). *The design and analysis of computer experiments*. Springer.

Schmidt P., Hartung J., Bennewitz J., & Piepho H. P. (2019). Heritability in plant breeding on a genotype-difference basis. *Genetics*, *212*, 991–1008. https://doi.org/10.1534/genetics.119.302134

Senn S. (2023). Student and the Lanarkshire milk experiment. *European Journal of Epidemiology*, *38*, 1–10. https://doi.org/10.1007/s10654-022-00941-x

Speed T. P. (1991). Introduction to Fisher (1926) The arrangement of field experiments. p.71–81. In: Kotz S., & Johnson N.L. (eds.) *Breakthroughs in Statistics*. *Volume II. Methodology and Distribution*. New York: Springer.

Stroup W. W. (2002). Power analysis based on spatial effects mixed models: A tool for comparing design and analysis strategies in the presence of spatial variability. *Journal of Agricultural Biological and Environmental Statistics*, *7*, 491–511. https://doi.org/10.1198/108571102780

Talbot M. (1984). Yield variability of crop varieties in the U.K. *Journal of Agricultural Science*, *102*, 315–321. https://doi.org/10.1017/S0021859600042635

Tedin O. (1931). The influence of systematic plot arrangement upon the estimate of error in field experiments. *Journal of Agricultural Science*, *21*, 191–208. https://doi.org/10.1017/S0021859600008613

von Lochow J., & Schuster W. (1961). *Anlage und Auswertung von Feldversuchen*: *Anleitungen und Beispiele für die Praxis der Versuchsarbeit*. Frankfurt: DLG-Verlag.

VSNi (2025). CycDesigN Version 8.0 http://www.vsni.co.uk/software/cycdesign

Williams, E. R., & Bailey, R. A. (2026). Extended semi-Latin squares for use in field and glasshouse trials. *Journal of Agricultural, Biological and Environmental Statistics* https://doi.org/10.1007/s13253-026-00729-6

Williams E. R., John J. A., & Whitaker D. (2014). Construction of more flexible and efficient p-rep designs. *Australian and New Zealand Journal of Statistics*, *56*, 89–96.

Williams E. R., & Piepho H. P. (2018). An evaluation of error variance bias in spatial designs. *Journal of Agricultural, Biological and Environmental Statistics*, *23*, 83–91. https://doi.org/10.1007/s13253-017-0309-2

Williams E. R., & Piepho H. P. (2019). Error variance bias in neighbour balance and evenness of distribution designs. *Australian and New Zealand Journal of Statistics*, *61*, 466–473. https://doi.org/10.1111/anzs.12277

Williams E. R., Piepho H. P., & Whitaker D. (2011). Augmented p-rep designs. *Biometrical Journal*, *53*, 19–27. https://doi.org/10.1002/bimj.201000102

Yates F. (1936). A new method of arranging variety trials involving a large number of varieties. *Journal of Agricultural Science*, *26*, 424–455. https://doi.org/10.1017/S0021859600022760

Yates F. (1939). The comparative advantages of systematic and randomized arrangements in agricultural and biological experiments. *Biometrika*, *30*, 440–466. https://doi.org/10.1093/biomet/30.3-4.440

Yates F. (1940). The recovery of inter-block information in balanced incomplete block designs. *Annals of Eugenics*, *10*, 317–325. https://doi.org/10.1111/j.1469-1809.1940.tb02257.x.

Yates F. (1981). *Sampling methods for censuses and surveys*. Charles Griffin.

Yates F., & Cochran W. G. (1938). The analysis of groups of experiments. *Journal of Agricultural Science*, *28*, 556–580. https://doi.org/10.1017/S0021859600050978.